\documentclass[aps,prl,twocolumn,groupedaddress, floatfix]{revtex4}

\usepackage{graphicx}
\usepackage{wasysym}

\def\Sg{{$\sigma_g$} }

\def\Pg{{$\pi_g$} }
\def\Pu{{$\pi_u$} }

\def\H2{H$_2$}
\def\N2{N$_2$}
\def\O2{O$_2$}
\def\CO2{CO$_2$}

\def\intensity[#1][#2]{$#1\times10^{#2}$~W/cm$^2$}

\begin{document}


\title{Alignment-Dependent Ionization of
       \N2, \O2, and \CO2 in Intense Laser Fields}

\author{Simon Petretti}

\author{Yulian V.~Vanne}

\author{Alejandro Saenz}

\affiliation{
     AG Moderne Optik,
     Institut f\"ur Physik,
     Humboldt-Universit\"at zu Berlin,
     Newtonstr.~15, D\,-\,12\,489 Berlin, Germany
}

\author{Alberto Castro}

\affiliation{
     Instituto de Biocomputaci\'on y F\'{\i}sica de Sistemas Complejos,
     Corona de Arag\'on 42, 
     50009 Zaragoza, Spain
}

\author{Piero Decleva}

\affiliation{
     Dipartimento di Scienze Chimiche,
     Universit{\'a} di Trieste,
     Via~L.~Giorgieri 1,
     I\,-\,34127 Trieste, Italy
}

\date{\today}



\begin{abstract}
     The ionization probability of \N2, \O2, and \CO2 in intense laser fields 
     is studied theoretically as a function of the alignment angle  
     by solving the time-dependent Schr\"odinger equation numerically assuming 
     only the single-active-electron approximation.
     The results are compared to recent experimental data 
     [D.~Pavi{\v{c}}i{\'c} et al., Phys.\,Rev.\,Lett.\ {\bf 98}, 243001
     (2007)] and good agreement is found
     for \N2 and \O2. For \CO2 a possible explanation is provided for the
     failure of simplified single-active-electron models to reproduce the 
     experimentally observed narrow ionization distribution. It is
     based on a field-induced coherent core-trapping effect. 
\end{abstract}


\maketitle

%
Time-resolved imaging of the dynamics of nuclei and electrons on a 
femtosecond or even sub-femtosecond time scale is a prerequisite for the 
experimental investigation of the formation and breaking of chemical bonds. 
Ultrashort laser pulses have recently been demonstrated to allow monitoring 
of nuclear motion with subfemtosecond and sub-{\AA}ngstrom  
resolution in real time \cite{sfm:goll06,sfm:ergl06,sfm:bake06}. It was 
also experimentally demonstrated that the high-harmonic radiation or electrons 
emitted in an intense laser pulse may in principle reveal electronic 
structure \cite{sfm:itat04,sfm:meck08} and thus have the potential for  
time-resolved imaging of changes of the electronic structure in, e.\,g., 
a chemical reaction. To reach this goal it is, however, important to 
understand the relation between electronic structure and strong-field 
response in molecules. 

In \cite{sfm:pavi07} the alignment dependence 
of the strong-field ionization was measured for \N2, \O2, and \CO2. 
The found relatively strong angular dependence of the ionization on the incident
electric field is interpreted in terms of molecular orbital theory: 
the measured ionization is supposed to map the structure of the highest 
occupied molecular orbital (HOMO). 
While the angle-resolved ionization of \N2 and \O2 reproduces the 
symmetry of the HOMO and was reasonably explained using a 
simplified strong-field model like molecular Ammosov-Delone-Krainov 
(MO-ADK) theory \cite{sfm:tong02}, this was not the case for \CO2. 

In order to find out why simple strong-field models like MO-ADK or the  
molecular strong-field approximation (MO-SFA) \cite{sfm:muth00} fail and 
are unable to properly reproduce the \CO2 results, 
a recently introduced approach (SAE-TDSE) to solve the time-dependent 
Schr\"odinger equation (TDSE) of a 
molecule within the single-active-electron approximation (SAE) 
\cite{sfm:awas08} is adopted in this Letter to calculate the alignment 
dependence 
of ionization of \N2, \O2, and \CO2. It avoids most of the assumptions  
made in the derivations of MO-ADK (a pure tunneling theory) or 
MO-SFA (that ignores the Coulomb interaction between the emitted electron 
and the remaining ion) 
which both neglect excited states. The reliability of the SAE-TDSE 
has been demonstrated, at least for H$_2$, in \cite{sfm:awas08} 
by a comparison to exact calculations. 
In this Letter we show that the failure of simple models to explain 
the experimental \CO2 results in \cite{sfm:pavi07} appears to be a 
laser-induced dynamical coherent coupling of the \mbox{HOMO} and 
the core, a phenomenon we name {\it coherent core trapping}.    

%

The SAE-TDSE approach \cite{sfm:awas08} and its present
extension to larger molecules and arbitrary orientation is only briefly 
described here.  
The basic assumption is that effectively only one single active
electron interacts with the external laser field and the corresponding
time-dependent Schr\"odinger equation is then solved numerically. 
The non-relativistic time-dependent Hamiltonian is treated semi-classically
and separated in an unperturbed time-independent molecular part $H_{\mathrm{mol}}$
and a time-dependent laser-electron interaction $H_{\mathrm{int}}(t)$.
A fixed molecular geometry is used to calculate the
field-free molecular electronic structure which is presently kept invariant
during the laser pulse. Experimental values of the equilibrium distances
have been adopted in the calculations. 
The time-dependent wavefunctions are expressed as a linear combination of 
the solutions of the time-independent field-free Hamiltonian
$H_{\mathrm{mol}}$. The latter are Kohn-Sham orbitals obtained from 
density-functional theory using the exchange correlation potential 
LB94 \cite{gen:vanl94}, since it yields the correct asymptotic behavior. 
The radial parts of 
the orbitals are expanded in a $B$-spline basis consisting of one set 
that is defined within one large central sphere (with radius 
$r_{\mathrm{max}}^0$) and additional sets defined within (non-overlapping) 
atom-centered spheres. While the latter sets assure a proper molecular 
description of the atomic cores, the former improves the description of 
molecular bonds and leads to a convenient ``box'' discretization of the 
electronic continuum. The angular parts are in both cases expressed in terms 
of spherical harmonics and molecular symmetry is fully accounted for. 
Converged results were obtained for radii $r_{\mathrm{max}}^0$ 
200.35~a.u., 160~a.u., and 200.35~a.u.\ and the values of $l_{\mathrm{max}}$  
10, 12, and 14 for \N2, \O2, and \CO2, respectively. 

In the original SAE scheme (method A) all but a single active electron are
frozen in the time propagation. The passive core orbitals are therefore 
removed from the active space in the TDSE propagation, i.\,e.\ the 
active electron is constrained to remain orthogonal to the {\it field-free}  
core. However, all electrons are driven by the field, and due to the  
unitary evolution under the time-dependent Hamiltonian all orbitals 
remain orthogonal at all times without the need of external orthogonality 
constraints. Thus in method B all orbitals are propagated within the 
complete functional space, including the one occupied by the  
{\it field-free} core. Due to the mentioned orthogonality, this does not 
violate the Pauli principle and insures in fact full gauge invariance 
of the results, as will be explained elsewhere. Mehod B may be dubbed 
independent-active-electron (IAE) approach and is in fact implicit in 
many model-potential calculations which, however, neglect orthogonality 
altogether.  

The inclusion of symmetries up
to $\Lambda=6$ where $\Lambda=0, 1, 2, \dots$ refers to $\sigma, \pi, \delta,
\dots$ symmetries was sufficient to reach convergence. The total 
number of states
used for time-propagation are limited by an energy cut-off parameter
which has been chosen between 10\,a.u.\ and 20\,a.u.\ in order to yield 
converged results.
This resulted in about 17,000, 22,000, and 22,000 states in the 
calculations for \N2, \O2, and \CO2, respectively.

The time profile of the applied linearly polarized laser pulse is represented 
by a cos$^2$ vector
potential and has in all cases shown a wavelength of 800\,nm and a duration of
40\,fs. While only results for the experimental peak 
intensities are shown, a number of calculations for lower peak intensities 
were performed and no qualitative changes were found for the total ion 
yields. Thus focal-volume 
averaging should not influence the main findings of this work, except one 
case as is discussed below.  
At the end of the time-propagation calculations (performed in length
gauge) the 
ionization yield is obtained from a sum over the populations of the discretized 
continuum states. In the case of H$_2$ it was demonstrated by a comparison to 
full two-electron calculations that for ionization yields not exceeding about 
10\,\% the SAE-TDSE results should be multiplied by a factor of 2 
in order to account for the two equivalent electrons \cite{sfm:awas08}. 
Correspondingly, the ionization yields obtained 
for a given initial orbital were multiplied with the occupation number and 
orbital degeneracy was explicitly considered. 
The total ionization is obtained by an 
incoherent sum of these weighted orbital yields. For the experimental 
parameters of \cite{sfm:pavi07} used in this work this sum remains well  
below 10\,\%, except for \O2 where the maxima almost reach 10\,\%. 
Since for \O2 the contributions from orbitals other than the HOMO are 
found to be negligible, this does, however, not change any of the conclusions 
of the present work.  

%

\begin{figure}
\begin{center}
    \includegraphics[width=0.45\textwidth]{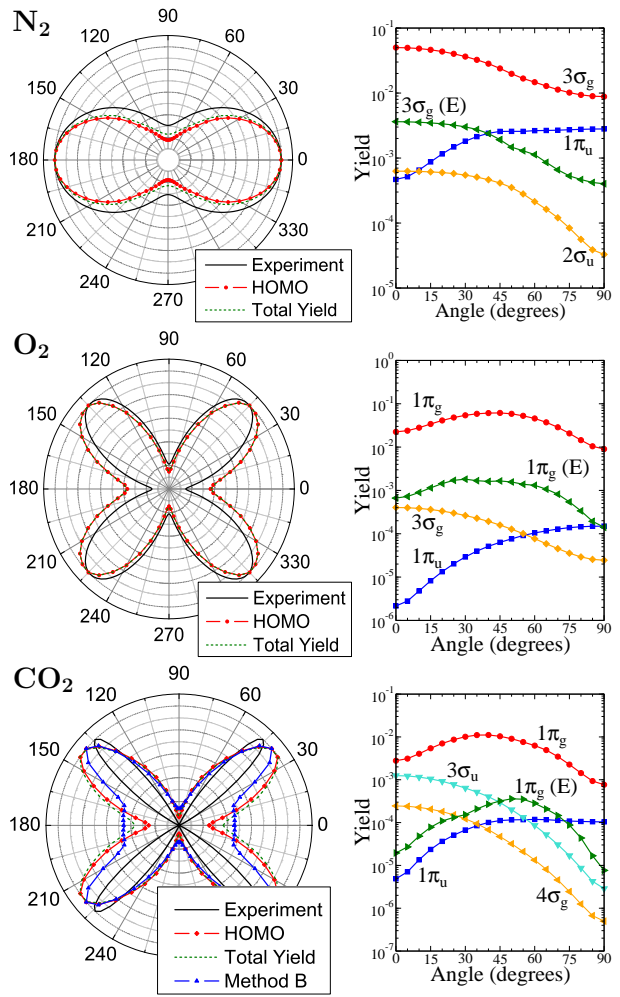}
    \caption{\label{fig:all_molecules}
    (Color online) {\bf Left panels:} Normalized ionization yields as a function
    of the angle between the electric field and the molecular axis: experimental data
    (black, \cite{sfm:pavi07}) and theoretical results (method A) for the 
    HOMO (red dots) and for the total yield (green dashed line). 
    For \CO2{} also the result of method B
    is shown. {\bf Right Panels:} Ionization
    yields of the HOMO and some lower lying orbitals (HOMO-X) as well as the 
    excitation yield (E) from the HOMO. { \bf Laser peak
    intensities:} \N2{} (\intensity[1.5][14]), \O2{} (\intensity[1.3][14]) and
    \CO2{} (\intensity[1.1][14]). 
    }
\end{center}
\end{figure}

Figure \ref{fig:all_molecules} shows a comparison of the orientation dependence
of the ionization yield of \N2 obtained in this work (method A) with the experimental
data in \cite{sfm:pavi07}. Already the pure HOMO ($3\,\sigma_g$) contribution 
agrees quite well with experiment, and this agreement is slightly improved, if the 
contributions of all orbitals are included. As the angle between molecular 
axis and field vector approaches 90$^\circ$, there is an increasing 
contribution from the \mbox{HOMO-1} (1$\pi_u$). Its inclusion changes the
ratio of parallel to 
perpendicular ionization from 5.7 to 4.4 which then is in reasonable agreement 
to the experimentally found 3.3$\pm 0.4$ \cite{sfm:pavi07}. This confirms 
the importance of the \mbox{HOMO-1} especially for perpendicular orientation,
also found recently for high harmonic generation in \cite{sfm:mcfa08}. 
The contribution of the 
\mbox{HOMO-2} ($2\sigma_u$) is on the other hand practically negligible, 
though at 0$^\circ$ it is larger than the one of the \mbox{HOMO-1}.   

For \O2 the agreement between SAE-TDSE and experiment is again  
good, as is seen from Fig.~\ref{fig:all_molecules}. The small shift of the 
maxima towards the horizontal 
axis may be explained by the influence of vibrational motion, 
since \O2 should preferentially ionize at smaller internuclear separations  
than at the equilibrium one \cite{sfm:saen00c}. 
The larger asymmetry between parallel and perpendicular orientation found in 
the SAE-TDSE is a consequence of the fact that both electrons 
in the degenerate \Pg orbitals ($\pi_g^x, \pi_g^y$) contribute equally 
in parallel orientation, 
but effectively only the $\pi_g^x$ electron contributes in $x$ direction. 
The symmetry observed in the experiment is thus somewhat surprising and 
it may be worthwhile to check the experimental deconvolution procedure 
in this respect. In fact, an even better agreement is found for all 
three considered molecules N$_2$, O$_2$, and CO$_2$, if the contributions 
from the $\pi^y$ orbitals are omitted.    
Compared with \N2, the alignment dependence of the ionization yield from the 
\O2 HOMO with its \Pg symmetry differs clearly, in accordance with 
expectations based on the corresponding 
(momentum) densities (see, e.\,g., Figs.\,2 and 3 in \cite{sfm:milo06}).   
The contributions from the \mbox{HOMO-1} ($3\sigma_g$) and \mbox{HOMO-2} 
($1\pi_u$) are found to be 
practically negligible, but their orbital structure is again 
imprinted in their ionization yields. For parallel alignment the 3\Sg orbital 
of \O2 is, e.\,g., much more likely to ionize than for perpendicular alignment and 
it shows a very similar behavior as the HOMO 3\Sg of \N2. The same analogy 
applies for the 1\Pu orbitals of \O2 and \N2.
The found correspondence between orbital structure and ionization behavior 
for HOMO and HOMO-1 of \N2 contradicts results of a recent all-electron 
time-dependent density-functional-theory calculation \cite{sfm:teln09}. 
Noteworthy, the present SAE-TDSE results for both \N2 and \O2 agree 
better to experiment than the results in \cite{sfm:teln09}. 

A similar correlation with the orbital shape is found for the orientation 
dependence of the HOMO and the \mbox{HOMO-X} (X=1, 2, 3) of \CO2. Although 
the HOMO of \O2 and \CO2 are both 1\Pg orbitals, the one of \CO2 is more 
stretched along the nuclear axis which leads to a more asymmetric 
orientation dependence (peaking at about 40$^\circ$). In that sense the 
present SAE-TDSE results confirm the conjecture that the alignment dependence 
of the strong-field ionization reflects the symmetry of the orbital. A 
prerequisite for this correspondence is the absence of intermediate resonances. 
The excitation yields to all electronically excited bound states are also shown in 
Fig.~\ref{fig:all_molecules}. For all three molecules 
they are much smaller than the ionization yield and, more importantly, follow 
in shape quite well the angular dependence of the HOMO ionization yields. 
Therefore, though excitation may not necessarily be negligible in absolute 
magnitude for the three considered molecules and laser parameters, 
it does not modify the orientation dependence of the ion yields.    

Nevertheless, the agreement of the \CO2 result with experiment is not good. The 
latter found very sharp maxima, the first peaking at about 46$^\circ$, and 
extremely small ionization in parallel and perpendicular orientation. 
Since the contribution from the \mbox{HOMO-2} ($3\sigma_u$) is non-negligible for the 
parallel orientation according to SAE-TDSE (in agreement with recent 
experimental evidence \cite{sfm:smir09}), the agreement gets even worse 
for the total ionization. Note, however, that the present results agree reasonably 
well with the measured angular distribution of fragments of \CO2 in 
\cite{sfm:alna05}. Furthermore, already in \cite{sfm:pavi07} the experimental 
\CO2 distribution was found to be disagreeing to simplified SAE models like 
MO-ADK and MO-SFA. Since the present SAE-TDSE is a stringent test of the SAE 
beyond such simplified models, it might be concluded that the deviation to 
experiment is a breakdown of the SAE itself, if it is not experimentally 
caused as is proposed in a recent compilation of theoretical results
\cite{sfm:zhao09}.  
The most likely explanation for the observed sharp angular distribution appears to be 
an intermediate resonance whose position is either badly or (in the case of 
doubly-excited states) not at all predicted within SAE. Clearly, an
experimental wavelength 
(or intensity) scan should reveal such a resonance and would thus be very 
desirable. In fact, there exists one further experimental investigation 
of the orientation dependent ionization of CO$_2$. In \cite{sfm:thom08} data 
are presented for an about 3 to 4 times lower peak intensity than in 
\cite{sfm:pavi07}. The distribution 
is much broader and thus in better agreement with simplified models.    

\begin{figure}
\begin{center}
     \includegraphics[width=0.45\textwidth]{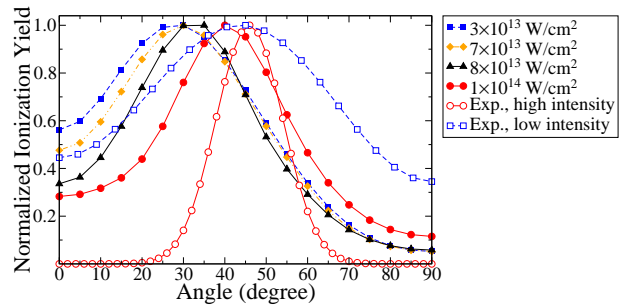}
     \caption{\label{fig:CO2_intensities}
     (Color online) Normalized ionization yield from the HOMO of \CO2 
     (method B) for different laser peak intensities between 
     \intensity[3][13] (as in the experiment
     \cite{sfm:thom08}) and \intensity[1.1][14] (experiment in
     \cite{sfm:pavi07}). 
     }
\end{center}
\end{figure}

So far, all discussed results were obtained with method A. While 
there is no qualitative change when applying method B to N$_2$ and O$_2$, 
this is not the case for CO$_2$ (see Fig.\,\ref{fig:all_molecules}). 
In this case the SAE-TDSE results are  
more similar to the experiment, since the peak shifts towards a larger 
angle and becomes narrower. Figure \ref{fig:CO2_intensities} shows the 
results obtained with method B for different laser intensities and reveals 
an even more interesting effect. The peak widths become pronouncedly narrower 
with increasing intensity. This is in accordance with the experimental 
results of \cite{sfm:pavi07} and \cite{sfm:thom08}. In fact, there is a 
pronounced change occurring at a peak intensity 
around \intensity[8][13] and thus in between the two experimentally
investigated intensities. A more detailed analysis reveals that this effect 
is due to a field-induced coupling between the field-free \mbox{HOMO} 
($1\pi_g$) and the \mbox{HOMO-1} ($1\pi_u$). In fact, at an intensity of 
about \intensity[8][13] the different ac Stark shifts of the two involved 
orbitals allow strong Rabi oscillations to occur and the resulting coherent 
trapping of the electron reduces the ionization probability. 

Due to symmetry, the $1\pi_g$ HOMO and the $1\pi_u$ \mbox{HOMO-1} are coupled 
by a field parallel to the molecular axis. Therefore, the effective intensity 
responsible for a coupling of these states is the component parallel to the 
molecular axis. For a given intensity the transition may be resonant for the 
parallel orientation, but is not yet resonant 
for other orientations. A further increase of the intensity allows then a 
larger range of orientations to become resonant. As a consequence, 
the orientational distribution becomes narrower, as the ionization probability 
decreases for an increasing range of angles close to 0$^\circ$ (or 180$^\circ$). 
It was confirmed that the shift of the maximum of the yield agrees with the 
corresponding shift of the projection of the laser field onto the molecular
axis. 

Note, there are two effects of the strong field that lead to coherent core 
trapping. First, the Stark shift brings the orbitals into resonance. (Note, 
however, that for N$_2$ the ac Stark shift suppresses the one-photon coupling 
between HOMO and HOMO-1 that in the field-free situation would be almost 
resonant for 800\,nm photons.) Second, 
the field-induced quiver motion temporarily depletes the {\it field-free} core 
orbital. Clearly, the present theoretical results alone would not be 
conclusive. Although the adopted approximation of an independent response 
of the electrons should be reasonable for intense fields 
(cf.\ \cite{sfm:awas08}), the results cannot be expected to be quantitatively 
correct. In fact, due to the intensity dependence also focal-volume effects are 
relevant. However, together with 
the experimental findings the present results strongly suggest that coherent 
core trapping is responsible for the unexpected experimental results observed 
for \CO2.  The field-free excited states (and their population) are on the
other hand not found to lead to substantial effects for the considered
examples, in contrast to the interpretation given in \cite{sfm:abus09}.

In summary, we have reproduced satisfactorily existing experimental data for
the angle-resolved ionization of the diatomic molecules \N2 and \O2 in intense 
short laser pulses. This shows the potential of the present SAE-TDSE approach that
had so far only been tested for \H2. These results support the principle idea of orbital 
mapping, but we also observe important contributions from the \mbox{HOMO-X}
that may distort the tomographical picture of the \mbox{HOMO} as in \N2. On
the other hand, \CO2 is an example for an intensity-dependent breakdown
of the orbital tomography that appears to be caused by a field-induced coupling of the 
\mbox{HOMO} with the core. This may explain why two experiments performed with 
different laser intensities observed different angular ionization patterns.

\begin{acknowledgments}
The authors would like to thank D.~Pavi{\v{c}}i{\'c} and I.~Thomann for kindly 
providing the experimental data and COST {\it CM0702} for financial support. 
SP, YV, and AS acknowledge financial support from the {\it Stifterverband 
f\"ur die Deutsche Wissenschaft}, the {\it Fonds der Chemischen Industrie},  
and {\it Deutsche Forschungsgemeinschaft} (SFB\,450/C6 and Sa936/2), 
AC within SFB\,658, and PD by CNR-INFM Democritos and INSTM Crimson.
\end{acknowledgments}



\end{document}